\documentclass[aps,showpacs,floatfix,twocolumn,pra,superscriptaddress]{revtex4-1}

\usepackage{amsmath}
\usepackage{graphicx}
\usepackage{bm,amssymb}
\usepackage{color,braket}
\usepackage{afterpage}

\begin{document}

\title{Berry Curvature of interacting bosons in a honeycomb lattice}

\author{Yun Li}
\affiliation{Centre for Quantum and Optical Science, Swinburne
  University of Technology, Melbourne, Victoria, 3122, Australia}

\author{Pinaki Sengupta}
\affiliation{School of Physical and Mathematical Science,
Nanyang Technological University, 21 Nayang Link, 637371 Singapore}
\affiliation{MajuLab, CNRS-UNS-NUS-NTU International
Joint Research Unit UMI 3654, Singapore}

\author{George G. Batrouni}
\affiliation{INLN, Universit\'{e} de Nice-Sophia Antipolis,
CNRS; 1361 route des Lucioles, 06560 Valbonne, France}
\affiliation{Institut Universitaire de France, 103 boulevard
  Saint Michel, 75005 Paris, France.}
\affiliation{Centre for Quantum Technologies, National
University of Singapore, 2 Science Drive 3, 117542 Singapore}
\affiliation{MajuLab, CNRS-UNS-NUS-NTU International
Joint Research Unit UMI 3654, Singapore}

\author{Christian Miniatura}
\affiliation{MajuLab, CNRS-UNS-NUS-NTU International
Joint Research Unit UMI 3654, Singapore}
\affiliation{Centre for Quantum Technologies, National
University of Singapore, 2 Science Drive 3, 117542 Singapore}
\affiliation{Department of Physics, National University of
Singapore, 2 Science Drive 3, 117542 Singapore}
\affiliation{INLN, Universit\'{e} de Nice-Sophia Antipolis,
CNRS; 1361 route des Lucioles, 06560 Valbonne, France}

\author{Beno\^{i}t Gr\'{e}maud}
\affiliation{MajuLab, CNRS-UNS-NUS-NTU International
Joint Research Unit UMI 3654, Singapore}
\affiliation{Centre for Quantum Technologies, National
University of Singapore, 2 Science Drive 3, 117542 Singapore}
\affiliation{Department of Physics, National University of
Singapore, 2 Science Drive 3, 117542 Singapore}
\affiliation{Laboratoire Kastler Brossel, UPMC-Sorbonne Universit\'es, 
CNRS, ENS-PSL Research University, Coll\`{e}ge de France, 4 Place Jussieu, 75005 Paris, France}

\begin{abstract}

  We consider soft-core bosons with onsite interaction loaded in the
  honeycomb lattice with different site energies for the two
  sublattices. Using both a mean-field approach and quantum
  Monte-Carlo simulations, we show that the topology of the honeycomb
  lattice results in a non-vanishing Berry curvature for the band
  structure of the single-particle excitations of the system. 
   This Berry curvature induces an anomalous Hall effect. It is seen by studying the 
   time evolution of a wavepacket, namely a superfluid ground state in a harmonic trap,
   subjected either to a constant force (Bloch oscillations) or to a sudden shift of the
   trap center.
\end{abstract}


\maketitle

\section{Introduction}

Topology and gauge fields, exemplified by Chern numbers, Berry phases,
Zak phase and Berry curvatures~\cite{Chern46,Berry84,Zak89}, are key
concepts at the heart of many condensed-matter
phenomena~\cite{TKNN82,Xiao10,Qi11}. For solid-state systems, Bloch's
theorem introduces wave numbers $\mathbf{k}$ belonging to a parameter
space with the topology of a torus, the Brillouin zone, and a set of
periodic wave functions depending parametrically on ${\bf k}$ which
offer natural settings for bundles and connections
\cite{Resta00}. Recently, ultracold atomic systems have progressively
and successfully confirmed their ability to mimic or emulate some
paradigmatic phenomena of condensed-matter systems, in particular
topological effects.  Indeed one can now load independent or
interacting bosons and/or fermions into two-dimensional, very well
controlled, optical lattices
\cite{ReviewBloch,Aidelsburger11,Stamperkurn12,Tarruell12,Uehlinger13}
and one can generate carefully designed synthetic magnetic
fields~\cite{Lin09, Cooper11a,Aidelsburger13,Ray14}.  These advances
pave the way to accurate cold atoms experiments targeting
topology-related effects such as defects~\cite{Gliga13}, color
superfluidity \cite{Rapp07}, momentum-space Berry
curvatures~\cite{Price12} or quantum Hall states with strong effective
magnetic fields~\cite{Jaksch03,Mueller04,Sorensen05,Cooper11b,
  Goldman13,Bloch13}.

Because of its remarkable low-energy electronic excitations, graphene
has been the source of many key
discoveries~\cite{Novoselov_04,Neto_09} which sparked a vivid research
flow now reaching new territories as exemplified by ultracold atoms
loaded in optical lattices~\cite{Zhu_07,GrapheneKL,
  Muramatsu10,Sengstock_11_a,Sengstock_11_b,
  Esslinger12,Esslinger13,Esslinger13b,Bloch13,krzysiek14}. In
particular, the tight-binding model on the honeycomb lattice or
equivalently on the brick wall lattice is well known to exhibit a
Berry curvature in its band
structure~\cite{Fuchs2010,Goldman13,keanloon14}, such that the
anomalous Hall effect can be observed~\cite{Xiao10,Carusotto04}. In
this paper, we discuss the impact on the Berry curvature of the
interactions between bosons loaded in the honeycomb lattice with site
energy imbalance, i.e. when the site energies of the two sub-lattices
are different. In that case, the phase diagram depicts Mott-insulating
phases at half-integer filling, a feature that is shared with the
energy imbalanced square lattice. On the other hand, as we show below,
the properties of the Bogoliubov excitations above the ground state
are different for these two lattices: only the excitations for the
honeycomb lattice depict a non-vanishing Berry curvature. This is
emphasized by looking at a dynamical situation, such as the Bloch
oscillations, where the anomalous Hall effect is observed.

The paper is organized as follows. In section~\ref{model}, we
introduce the model, summarize the basic properties of the energy
imbalanced honeycomb lattice, in particular the Berry curvature and we
present the two approaches used in the paper: the quantum Monte-Carlo
method (QMC)~\cite{Sandvik1991,Sandvik1997,Sandvik1999,Sengupta2005}
and the Gutzwiller ansatz
(GA)~\cite{tcr,ageorge_varenna,tutorialLew}. In section~\ref{results},
we present our numerical results: the ground state phase diagram and
the excitations, emphasizing that they depict non-vanishing Berry
curvature. In section~\ref{dynamics}, we explain how to observe the
anomalous Hall effect resulting from the Berry curvature in the
excitations. A summary of results and conclusions are given in
section~\ref{conclusion}.

\section{Model}
\label{model}
\subsection{Single-particle Hamiltonian and Berry Curvature}

\begin{figure}[h!]
\includegraphics[scale=0.5]{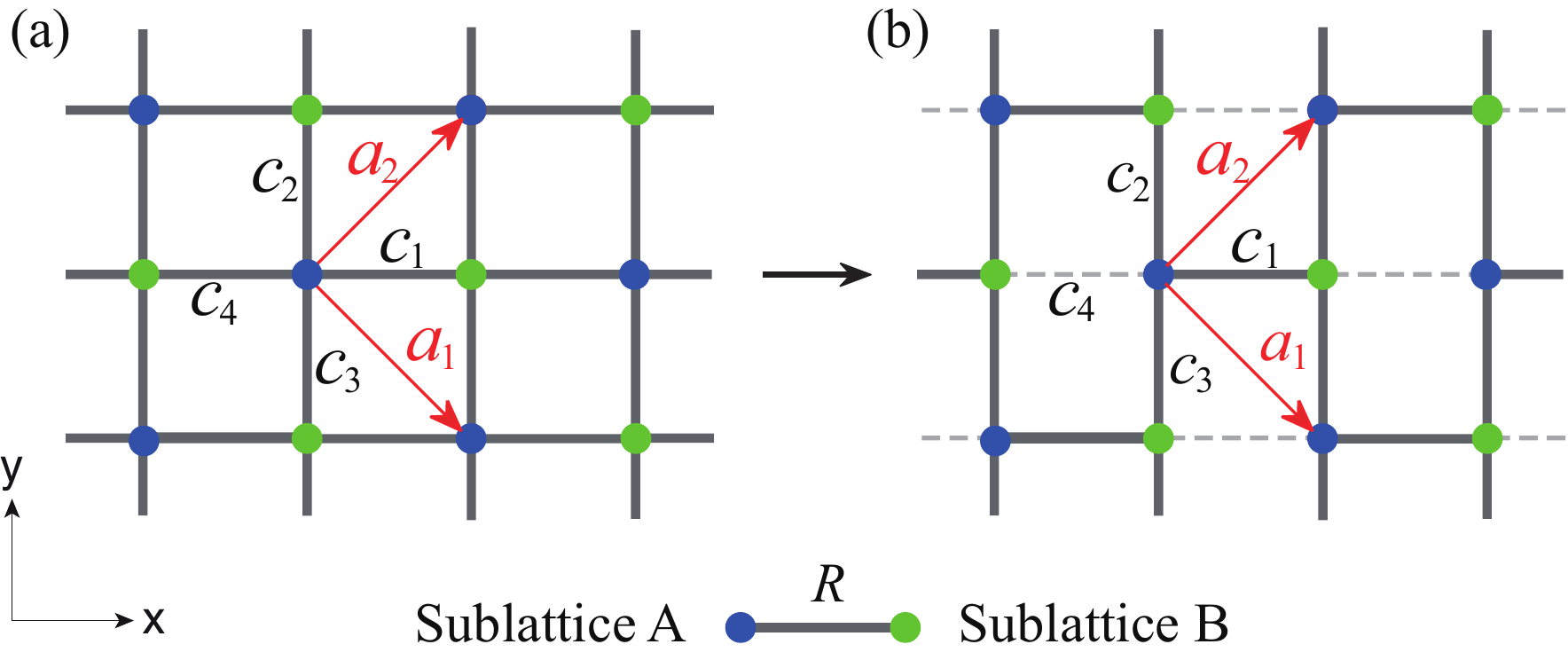}
\caption{(Color online) Structure of a bipartite lattice with four nearest neighbors.
  Sites belonging to sublattice $A$ are represented by dark (blue)
  points in the figure, whereas $B$ sites are represented by light grey (green)
  points. The hopping amplitudes are all equal
  in (a) while in (b)  the hopping term $t_4$ along links $\pmb{c}_4$ is different from the other three near
  neighbor hopping terms.} \label{fig:lattice}
\end{figure}

We start with a general bipartite lattice structure with four nearest
neighbors. Each primitive unit cell contains two lattice sites $A$ and
$B$, as depicted in Fig.~\ref{fig:lattice}. The whole lattice
structure can be generated by repeated translations of a unit cell
along the Bravais primitive vectors
\begin{equation}
  \pmb{a}_1 = \hat{\pmb{e}}_x - \hat{\pmb{e}}_y \quad \text{and} \quad
  \pmb{a}_2 = \hat{\pmb{e}}_x + \hat{\pmb{e}}_y
\end{equation}
where we have assumed the spacing between two sites in the primitive
unit cell is unity. The primitive reciprocal lattice vectors are
\begin{equation}
\pmb{b}_1 = \pi(\hat{\pmb{e}}_x - \hat{\pmb{e}}_y) \quad \text{and}
\quad \pmb{b}_2 = \pi(\hat{\pmb{e}}_x +\hat{\pmb{e}}_y),
\end{equation}
and fulfills  the
relation $\pmb{a}_i \cdot \pmb{b}_j=2\pi\delta_{ij}$.

The displacements that move a $A$ site to a neighboring $B$ site are
parameterized by bond vectors
\begin{equation}
\begin{aligned}
  \pmb{c}_1 & =-\pmb{c}_4 = \hat{\pmb{e}}_x \\
  \pmb{c}_2 &=-\pmb{c}_3 = \hat{\pmb{e}}_y
\end{aligned}
\end{equation}

Atoms trapped in the lattice can hop from one site to the neighboring
sites with amplitudes $t_\nu$ ($\nu=1\, ..\,4$) corresponding to the
four bond vectors. In the absence of interactions, the Hamiltonian in
real space reads
\begin{equation}
 H_0 = -\sum_{\langle i,j\rangle}\left(t_{ij} \,b_i^\dag b_j +
     t_{ij}^\ast \,b_j^\dag b_i \right)-\mu_A\sum_{i\in A} n_i -
   \mu_B \sum_{j\in B} n_j \label{eq:Ham_r}
\end{equation}
where $t_{ij} \in \{t_1,\, t_2,\, t_3,\, t_4\}$, $b_i$ is the
operator annihilating one particle on lattice site $i$, $n_i =
b^\dag_i b^{}_i$ is the atom number operator, $\mu_A$ and
$\mu_B$ are chemical potentials for sublattice $A$ and $B$
respectively. When all hopping amplitudes $t_\nu$ are the same,
Eq.~\eqref{eq:Ham_r} describes a standard square lattice. If $t_4=0$,
the lattice becomes the brick wall lattice which is topologically
equivalent to the honeycomb lattice. Using Fourier transform, we
obtain the Hamiltonian in momentum space
\begin{equation}
  H_0 = \sum_{\mathbf{k}}\begin{pmatrix}b_{\mathbf{k}A}^\dag &
    b_{\mathbf{k}B}^\dag\end{pmatrix}\begin{pmatrix}\Delta &
    \Gamma^{}_{\mathbf{k}} \\ \Gamma^{\ast}_{\mathbf{k}} & -\Delta
  \end{pmatrix} \begin{pmatrix} b_{\mathbf{k}A} \\
    b_{\mathbf{k}B}\end{pmatrix}- \mu N \label{eq:Ham_k}
\end{equation}
where $\mathbf{k}$ is the Bloch wave vector, $\Delta = -(\mu_A -
\mu_B)/2$, $\mu=(\mu_A+\mu_B)/2$, $N= \sum_{\mathbf{k}}
(b_{\mathbf{k}A}^\dag b_{\mathbf{k}A}+ b_{\mathbf{k}B}^\dag
b_{\mathbf{k}B})$ is the total number of particles, and
$\Gamma_{\mathbf{k}} = -\sum_{\nu = 1}^4 t_\nu \, e^{-i \mathbf{k}
  \cdot \pmb{c}_\nu}$. The Hamiltonian \eqref{eq:Ham_k} has the
eigenvalues
\begin{equation}
\epsilon_{\pm}(\mathbf{k}) = \pm \sqrt{\Delta^2+ |
\Gamma_{\mathbf{k}} |^2} -\mu
\end{equation}
and eigenfunctions \cite{Fuchs2010}
\begin{equation}
\label{eq:eigenfunctions_k}
|\psi_{\mathbf{k}+}\rangle =\begin{pmatrix}\cos \dfrac{\beta}{2} \\
  \sin\dfrac{\beta}{2} e^{i\phi}\end{pmatrix}, \quad
|\psi_{\mathbf{k}-}\rangle =\begin{pmatrix} -\sin\dfrac{\beta}{2}
  e^{-i\phi} \\ \cos\dfrac{\beta}{2}\end{pmatrix}
\end{equation}
where $\phi = -\arg \Gamma_{\mathbf{k}}$ and $\beta = \arcsin
(\,|\Gamma_{\mathbf{k}}| / \sqrt{\Delta^2+ |\Gamma_{\mathbf{k}}|^2}
\,)$. For $\Delta=0$, the gap between the two energy bands closes at
$|\Gamma_{\mathbf{k}}|=0$, and exhibit the so-called Dirac cone. To open a gap, one
needs to imbalance the chemical potential between the $A$ and $B$
sites. In the following, we always consider the gapped case,
i.e. $\Delta \neq 0$.

The Berry curvature for each band of the Hamiltonian~\eqref{eq:Ham_k} 
can be calculated as \cite{Fuchs2010}
\begin{equation}
  \Omega_{\pm}(\mathbf{k}) = \pm\frac{1}{2}\left(
    \frac{\partial\cos\beta}{\partial k_x}\frac{\partial\phi}{\partial k_y}-
    \frac{\partial\cos\beta}{\partial k_y}\frac{\partial\phi}{\partial k_x}\right)
        \label{eq:Berry_EPJB}
\end{equation}
 Expanding the
Bloch wave vector $\mathbf{k}$ over the reciprocal primitive vectors,
$\mathbf{k}=\alpha_1 \pmb{b}_1 +\alpha_2 \pmb{b}_2$,  we find
the Berry curvature \eqref{eq:Berry_EPJB} can be written as follows
\begin{equation}
\label{eq:Berry_real_t}
\begin{aligned}
  \Omega_{\pm}(\mathbf{k}) &= \pm \dfrac{1}{2}
  \dfrac{\Delta}{\left(\Delta^2+|\Gamma_{\mathbf{k }}|^2 \right)^{3/2}}\\
  \times&\bigg\{\left[t_1 t_3 - t_2 t_4\right]\sin(2\pi\alpha_2)-\left[t_1 t_2 - t_3 t_4\right]\sin\left(2\pi\alpha_1
  \right)\bigg\}
\end{aligned}
\end{equation}
where we have assumed real values for the hopping amplitudes. 
For the square lattice ($t_\nu = t$), the Berry curvature is always zero, 
reflecting the presence of both the inversion and the time-reversal symmetry in the system. 
However, if one of the hopping amplitudes is different from the other three, 
for example $t_4 \ne t_{1,2,3}$, the symmetry under the exchange $x \to -x$ is broken. 
As a consequence, the system can exhibit a non-vanishing Berry curvature. 
In this case, as the symmetry under the exchange $y \to -y$ remains, 
$\beta({\mathbf{k}})$ and $\phi(\mathbf{k})$ are even functions of $k_y$. 
According to Eq.~\eqref{eq:Berry_EPJB}, one can easily see that the Berry curvature should be an odd function of 
$k_y$, i.e. $\Omega_\pm(k_x,k_y) = -\Omega_\pm(k_x,-k_y)$. 

\begin{figure}[h!]
\includegraphics[scale=0.5]{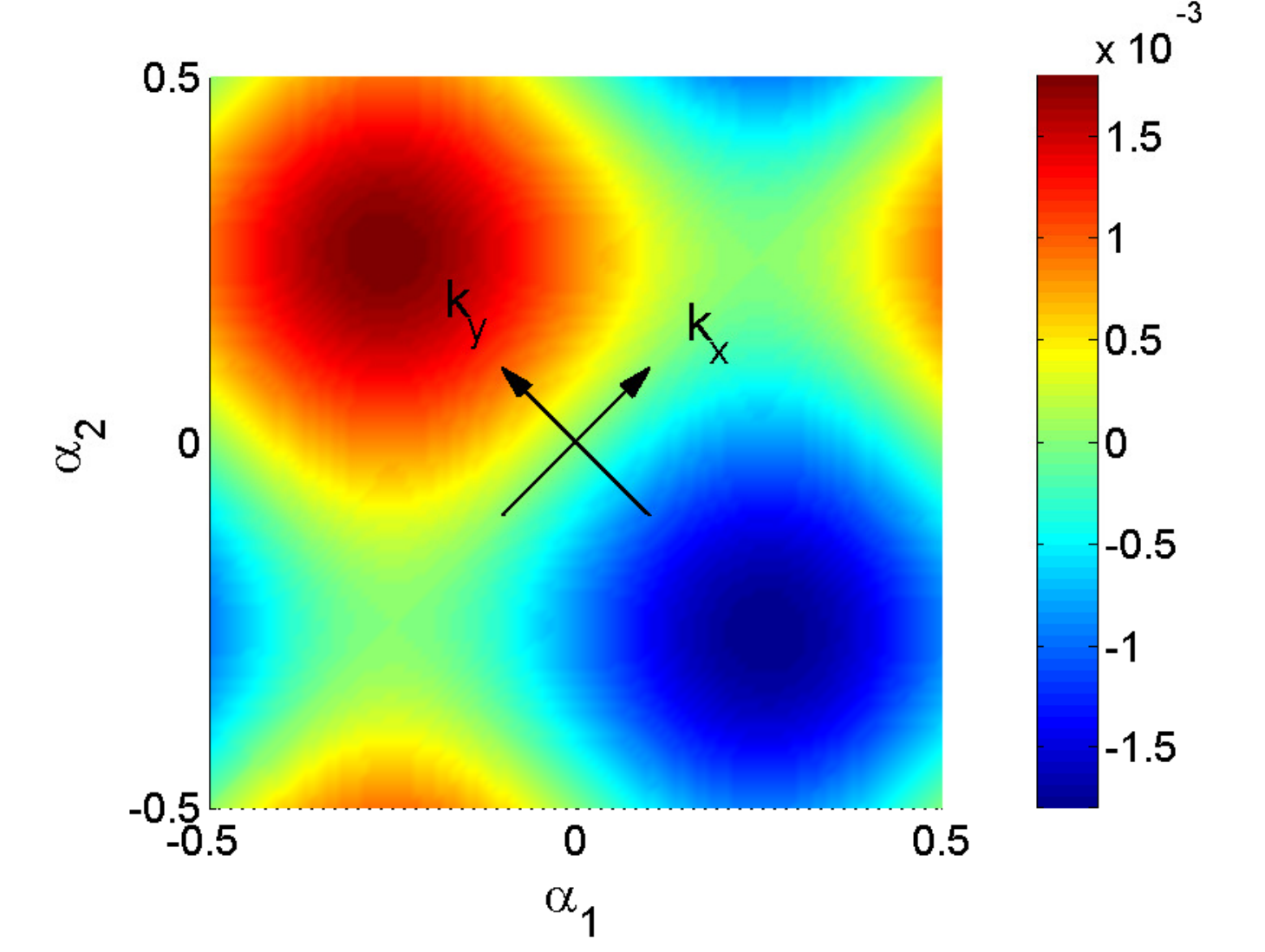}
\caption{(Color online) Berry curvature in the first Brillouin zone
  for the honeycomb lattice calculated from
  Eq. \eqref{eq:Berry_real_t}. The parameter are chosen as follows:
  $t_{1,\,2,\,3}=t$, $t_4=0$, $\Delta=5t$. }
\label{fig:Berry_single}
\end{figure}

In Fig.~\ref{fig:Berry_single}, we give an example of the Berry curvature in the first Brillouin zone ($\mathrm{BZ}$),
i.e., for $|\alpha_i|<1/2$,  calculated from
\eqref{eq:Berry_real_t}
for the imbalanced honeycomb lattice, i.e. $t_{1,2,3} = t$, $t_4 = 0$ and $\Delta=5t$. Due to the time-reversal symmetry, the Berry curvature 
has the usual property $\Omega_\pm(-\mathbf{k}) = -\Omega_\pm(\mathbf{k})$ such that, for each band, the Chern number 
which is the surface integral of Berry curvature, $\gamma_\pm = \int_{\mathrm{BZ}} \text{d}S\, \Omega_\pm(\mathbf{k})$, is vanishing. 
This emphasizes that the bands are topologically trivial. Furthermore, due to the odd symmetry of the Berry curvature under the 
exchange $k_y \to -k_y$, it is also an even function of $k_x$, $\Omega_\pm(k_x,k_y) = \Omega_\pm(-k_x,k_y)$, as shown in Fig.~\ref{fig:Berry_single}: 
in the $(\alpha_1,\alpha_2)$ plane, the Berry curvature is symmetric with respect to the anti-diagonal and antisymmetric with respect to the diagonal.

\subsection{Numerical methods}

When the interaction on a site is taken into account, the system can
be described by the Bose-Hubbard Hamiltonian
\begin{equation}
H= H_0 +\dfrac{U}{2}\sum_i n_i \left(n_i - 1\right)
\label{eq:Ham_BH}
\end{equation}
where $H_0$ is given by \eqref{eq:Ham_r}, and $U$ is the on-site
interaction strength. This is the Hamiltonian we shall investigate in
this study using two complementary methods that are described below.

\subsubsection{Quantum Monte-Carlo}

The ground state of Hamiltonian \eqref{eq:Ham_BH} can be studied using
the stochastic series expansion (SSE) quantum Monte Carlo method with
operator-loop updates \cite{Sandvik1991,Sandvik1997,Sandvik1999}.  The
SSE is a finite-temperature QMC algorithm based on importance sampling
of the diagonal matrix elements of the density matrix $e^{-\beta {\cal
    H}}$ .  There are no approximations beyond statistical
errors. Using the ‘‘operator-loop’’ cluster update, the
autocorrelation time for the system sizes we consider here is at most
a few Monte Carlo sweeps for the entire range of parameter space
explored.  The simulations are carried out on finite lattices with
$L^2$ sites for $L$ up to 32 at temperatures sufficiently low in
order to resolve ground-state properties of this finite system
\cite{Sengupta2005}.  Estimates of physical observables in the
thermodynamic limit are obtained from simultaneous finite-size and
finite-temperature extrapolation to the $L\to \infty, \beta\to \infty$
limit.  We consider here the soft-core boson case which allows
multiple occupation  to occur (up to $n^{(\max)}=4$ bosons per lattice site are allowed in
the present study). Periodic boundary conditions along the $x$ and $y$
axis have been used.

\subsubsection{Mean-field}

A well known mean-field method to solve the Bose-Hubbard model is the
Gutzwiller ansatz~\cite{Jaksch98,Sachdev,bosonsGeorge,lewenstein07},
where the ground state wavefunction is assumed to be a tensor product
of on-site wavefunctions:
\begin{equation}
|\Psi \rangle = \bigotimes_j |\psi_j\rangle\text{ where }
|\psi_j\rangle = \sum_{n=0}^{N} f_{n,j} |n,j\rangle
\label{eq:Gutzwiller}
\end{equation}
where $|n,j\rangle$ represents the Fock state of $n$ atoms occupying
the site $j$, $N$ is a cutoff in the maximum number of atoms
per site, and $f_{n,j}$ is the probability amplitude of having the
site $j$ occupied by $n$ atoms.

Minimizing the mean-field energy $\langle \Psi| H |\Psi\rangle$ over
the amplitudes $f_{n,j}$ allows us to determine the mean-field ground
state properties as functions of the different parameters $(U,\,t,\,
\mu_A, \, \mu_B)$. For instance, the superfluid phase corresponds to a
non-vanishing value of the order parameter 
$\langle\Psi|b_i|\Psi\rangle$, whereas the Mott
phase corresponds to a vanishing order parameter and the $|\psi_i\rangle$ are
pure Fock states.

In addition, the preceding ansatz~\eqref{eq:Gutzwiller} can be
extended to the time domain giving us access not only to the
Bogoliubov excitations above the ground state, but also to the full
mean-field evolution of the interacting wavefunction. More precisely,
the time evolution is obtained by solving the following set of
equations:
\begin{equation}
 \label{eq:mf_dynamics}
 i\frac{\text{d} f_{n,j}(t)}{\text{d}t}=\frac{\partial \langle
   \Psi(t)| H |\Psi(t)\rangle}{\partial f_{n,j}^\ast}.
\end{equation}

The Bogoliubov excitations are obtained by expanding the amplitudes
$f_{n,j}(t)$ around their ground state values $f^{(0)}_{n,j}$, namely:
\begin{equation}
f_{n,j} = \left[ f^{(0)}_{n,j}+g_{n,j}(t)\right] e^{-i \omega_j t}.
\end{equation}
where $\omega_j$ is the frequency of the ground state evolution.
Assuming $|g_{n,j}(t)| \ll |f^{(0)}_{n,j}|$ and keeping only the
linear terms in the dynamical Eqs.~\eqref{eq:mf_dynamics}, one obtains
the usual Bogoliubov equations:
\begin{equation}
\label{eq:bogo_eq}
 i\frac{\text{d}}{\text{d} t} \begin{bmatrix} g(t)\\  g^\ast(t)
 \end{bmatrix} =  \mathcal{L} \begin{bmatrix} g(t)\\ g^\ast(t)
 \end{bmatrix},
\end{equation}
where $g(t)$ is a shorthand notation for the vector $(\cdots, \,
g_{0,j}, \, g_{1,j}, \, \cdots, \, g_{N,j}, \, \cdots)^T$ and
$\mathcal{L}$ has the usual Bogoliubov structure:
\begin{equation}
 \mathcal{L} = \begin{pmatrix} \mathcal{A} & \mathcal{B} \\ -\mathcal{B}^\ast & -\mathcal{A}^\ast
 \end{pmatrix}.
\end{equation}
$\mathcal{A}$ and $\mathcal{B}$ are complex matrices satisfying $\mathcal{A}^{\dagger}=\mathcal{A}$ and
$\mathcal{B}^T=\mathcal{B}$.

Due to the $U(1)$ invariance, the values of $f^{(0)}_{n,j}$ can be
taken real. Furthermore, as we have observed (see below), the
mean-field ground state does not break the translation invariance,
such that the values $f^{(0)}_{n,j}$ are the same for all
\textit{equivalent} sites, namely $f^{(0)}_{n,j} = A_{n}$ for all
$A-$sites and $f^{(0)}_{n,j}= B_{n}$ for all $B-$sites. Because of
translation invariance, the Bogoliubov Eqs.~\eqref{eq:bogo_eq} can be
diagonalized in momentum space, taking into account the bipartite
nature of the lattice:
\begin{eqnarray}
  g_{n,j}(t)& = & \sum_{\mathbf{k}} e^{i(\mathbf{k} \cdot \mathbf{A}_j-
    \omega t)} u_{\mathbf{k},A,n} + e^{-i(\mathbf{k} \cdot \mathbf{A}_j -
    \omega t)} v^*_{\mathbf{k},A,n} \quad \\
  g_{n,j}(t)& = & \sum_{\mathbf{k}} e^{i(\mathbf{k} \cdot \mathbf{B}_j-
    \omega t)} u_{\mathbf{k},B,n} + e^{-i(\mathbf{k} \cdot \mathbf{B}_j-
    \omega t)} v^*_{\mathbf{k},B,n} \quad
\end{eqnarray}
for $A$ sites and $B$ sites respectively, leading to the following
eigensystem
\begin{equation}
 \omega_{\mathbf{k}} \begin{pmatrix} u_{\mathbf{k}} \\ v_{\mathbf{k}}
 \end{pmatrix} = \mathcal{L}_{\mathbf{k}} \begin{pmatrix}
 u_{\mathbf{k}} \\ v_{\mathbf{k}} \end{pmatrix}
\end{equation}
where 
$u_{\mathbf{k}}=(u_{\mathbf{k},A,0},\cdots,u_{\mathbf{k},A,N},u_{\mathbf{k},B,0},\cdots,u_{\mathbf{k},B, N})$
and $v_{\mathbf{k}} =(v_{\mathbf{k},A,0}, \cdots,v_{\mathbf{k},A,N}, v_{\mathbf{k},B,0},\cdots,v_{\mathbf{k},B,N})$.
$\mathcal{L}_{\mathbf{k}}$ has the same structure as $\mathcal{L}$,
which leads to the usual property: if $(u,v)$ is an eigenstate for the
energy $\omega$, then $(v^\ast,u^\ast)$ is an eigenstate for the
energy $-\omega^\ast$. Therefore one can focus on the eigenstates with
positive skew-norm only, i.e. such that $u^\dag u- v^\dag v = 1$. The
ground state is stable if all these eigenstates have real and positive
eigenenergies $\omega_{\mathbf{k}}$.

\section{Results}
\label{results}
\subsection{Phase diagram}

\begin{figure}[h!]
\includegraphics[scale=0.6]{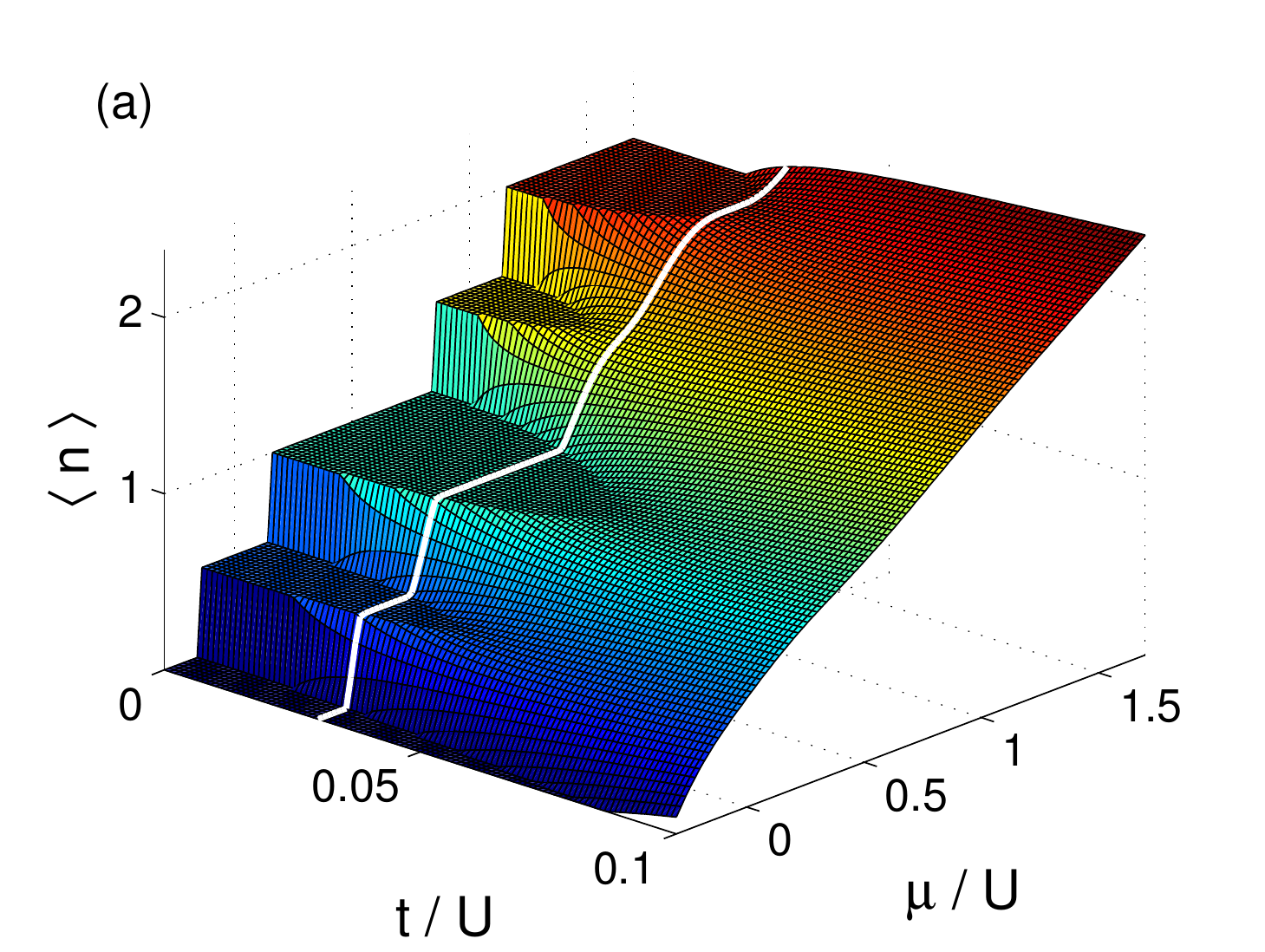}
\includegraphics[scale=0.5]{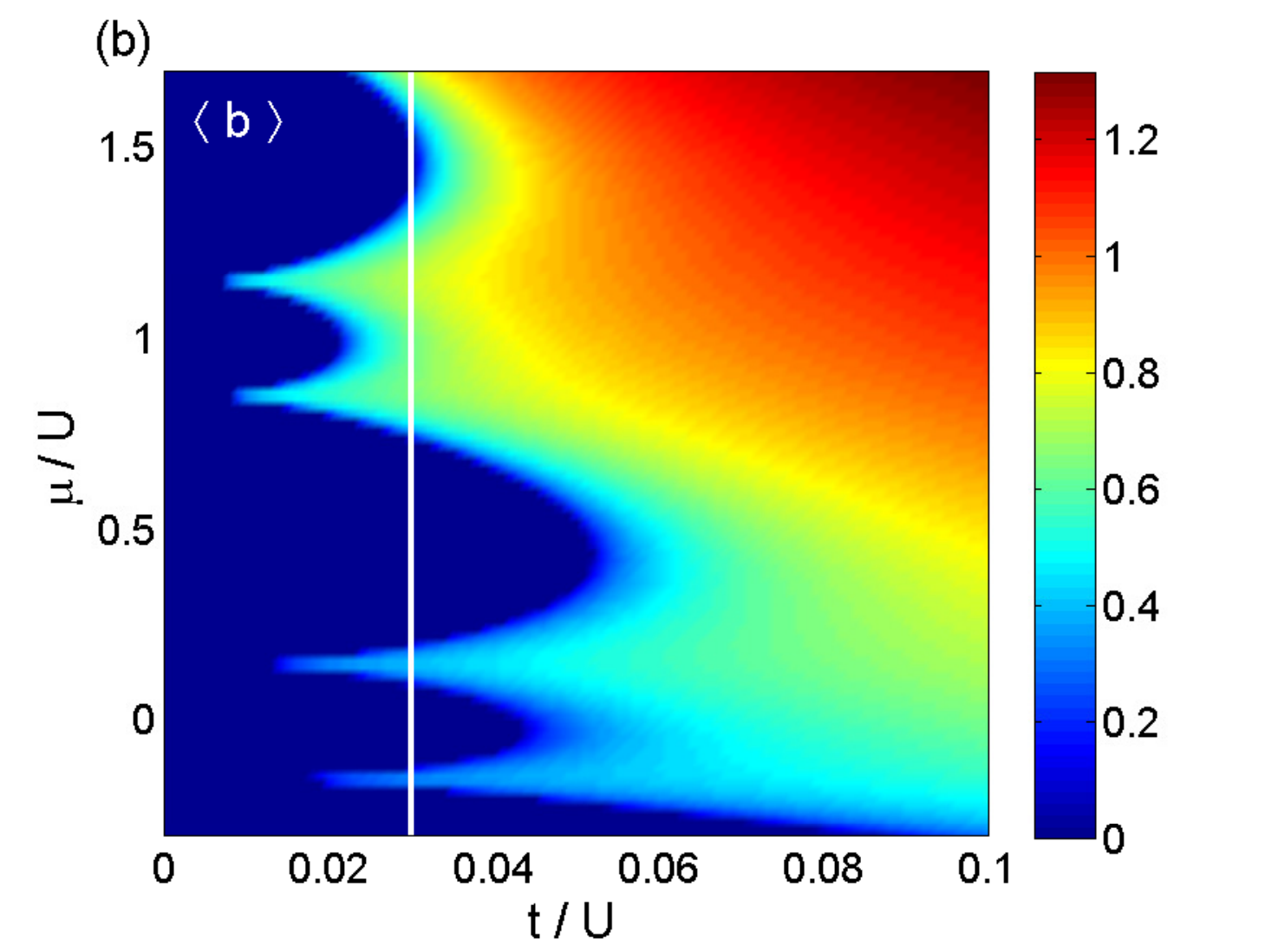}
\caption{(Color online) Mean-field phase diagram of the Bose-Hubbard
  Hamiltonian Eq. \eqref{eq:Ham_BH} on the honeycomb lattice. (a)
  Density per site; (b) the superfluid order parameter. The parameters
  used in the calculation are $t_{1,\,2,\,3}=t$, $t_4=0$, $\Delta/U
  =0.15$, $N = 10$. The white lines correspond to $t/U =
  0.03$. }
\label{fig:phase_MF}
\end{figure}

\begin{figure}[h!]
\includegraphics[scale=0.55]{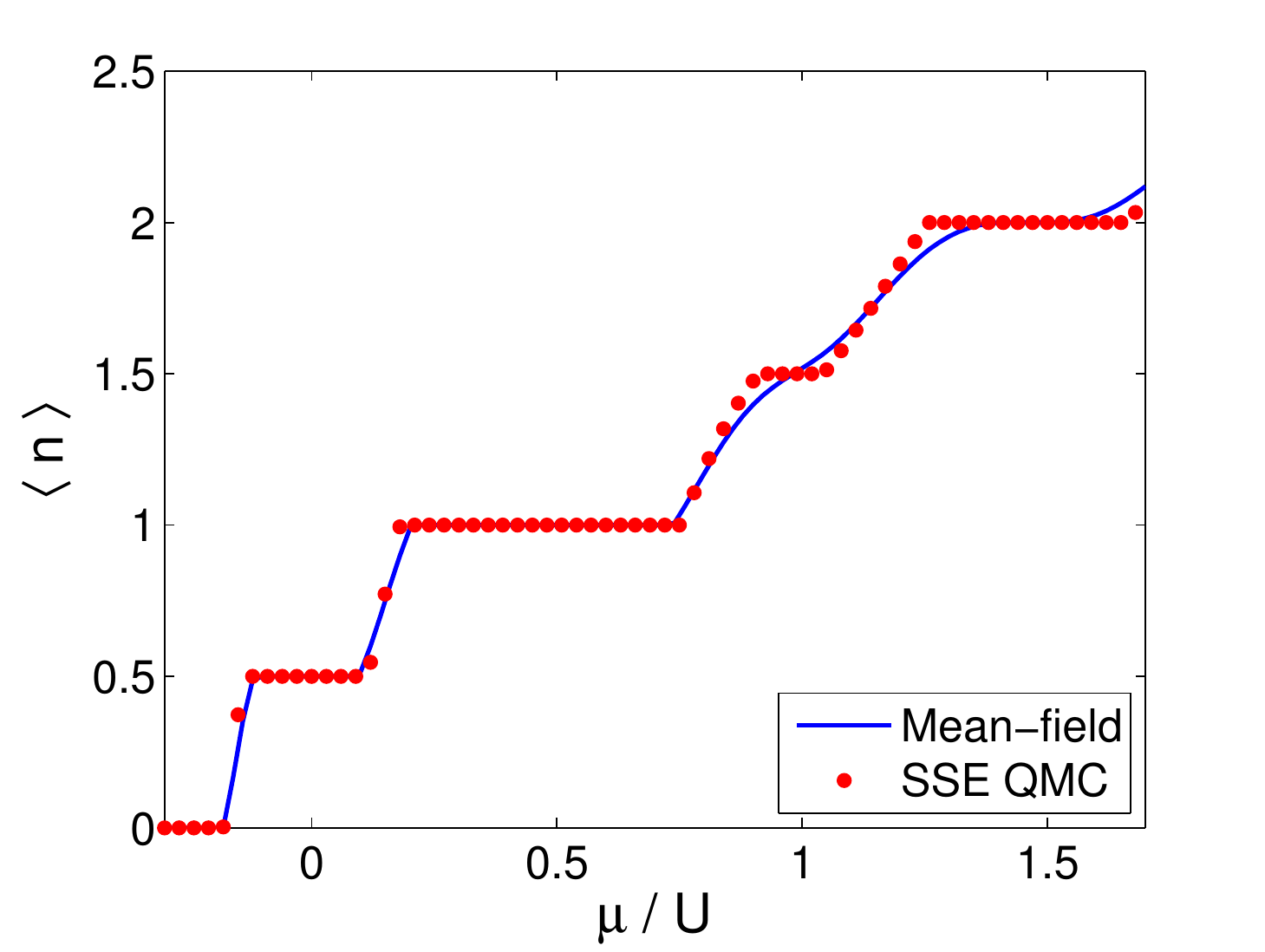}
\caption{(Color online) Comparison of QMC and mean-field results for the density per
  site as a function of total chemical potential
  $\mu=(\mu_A+\mu_B)/2$. Parameters are $t_{1,\,2,\,3}=t$, $t_4=0$,
  $\Delta/U=0.15$, as in Fig.~\ref{fig:phase_MF}, and the hopping
  amplitude is fixed to $t/U=0.03$. The blue solid curve is obtained
  from the mean-field approach (corresponding to the white line in
  Fig.~\ref{fig:phase_MF}), the red dots are obtained from QMC
  simulations. The lattice size for QMC simulations is $ 16\times 16$,
  with a maximum atom number per lattice site
  $n^{(\max)}=4$. } \label{fig:phase_QMC}
\end{figure}

Figure \ref{fig:phase_MF} shows the ground state phase diagram for a
honeycomb lattice ($t_4=0$) obtained from ansatz
Eq.(\ref{eq:Gutzwiller}).  Due to different chemical potentials
between the $A$ and $B$ sites, the insulating lobes occur not only at
$\langle n\rangle= 0,\,1,\,2, \,\ldots$ but also at $\langle n\rangle
=1/2,\,3/2,\,\ldots$. We compare the mean-field result with QMC
simulation in Fig.~\ref{fig:phase_QMC}, where we plot the density per
site as a function of $\mu=(\mu_A+\mu_B)/2$, fixing the hopping
amplitude at $t/U=0.03$, which corresponds to the white lines in
Fig.~\ref{fig:phase_MF}. Very good agreement is found between the
mean-field approach and QMC simulations.

\subsection{Excitations}

The phase diagram in Fig.~\ref{fig:phase_MF} itself is not a signature
of the Berry curvature induced by  the topology of the honeycomb
lattice. Indeed, the Hubbard model on the square lattice (i.e.  with a
non-vanishing hopping parameter $t_4$) with the same imbalance
$\mu_A-\mu_B$ exhibits a similar phase diagram, including the
half-integer filling Mott phases. The reason is that the ground state
corresponds, roughly speaking, to $\mathbf{k=0}$, precisely where the
Berry curvature $\Omega(\mathbf{k})$ vanishes. Therefore, the impact
of the Berry curvature has to be found in the excitation properties.

\begin{figure}[h!]
\includegraphics[scale=0.55]{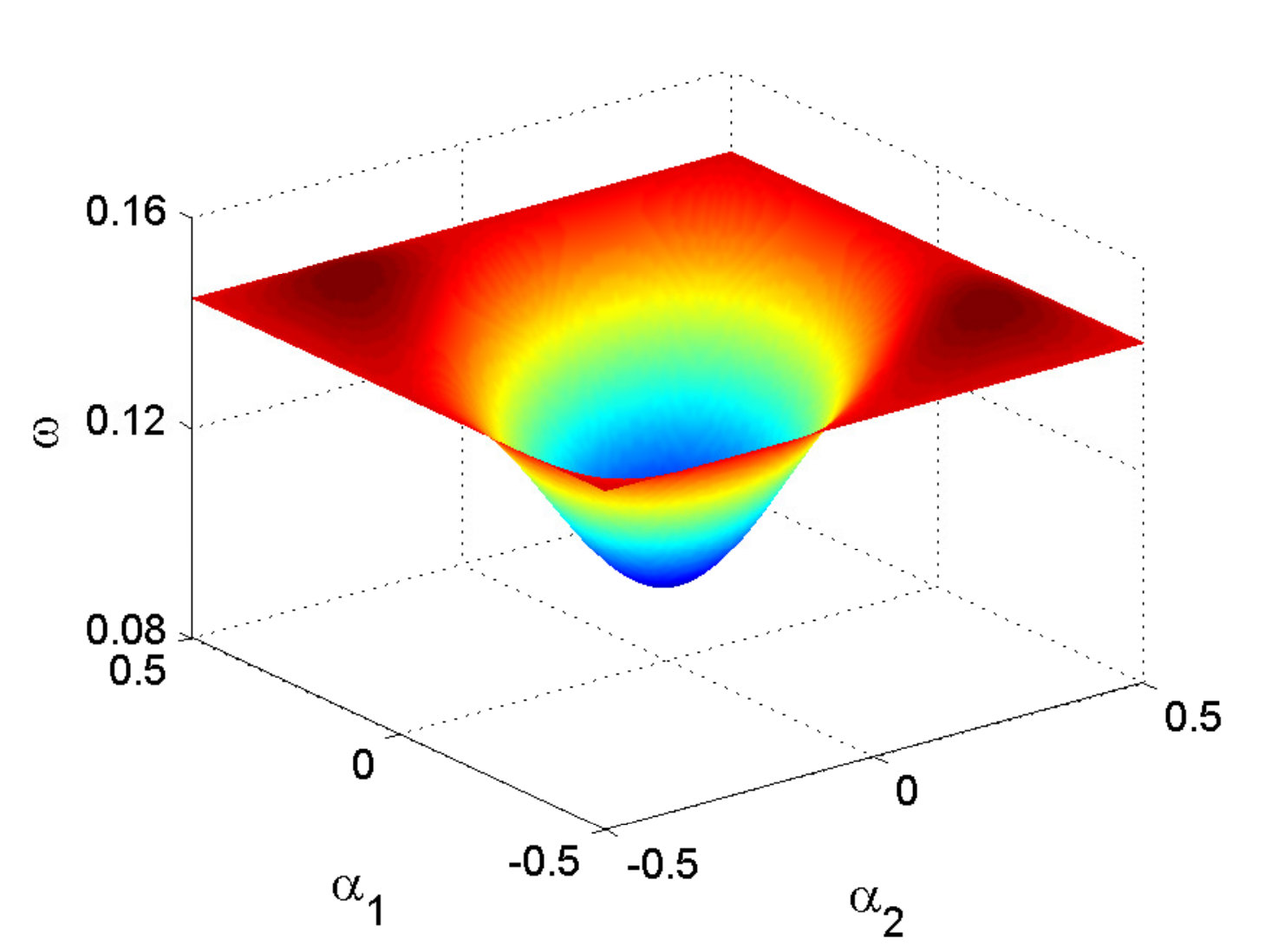}
\includegraphics[scale=0.55]{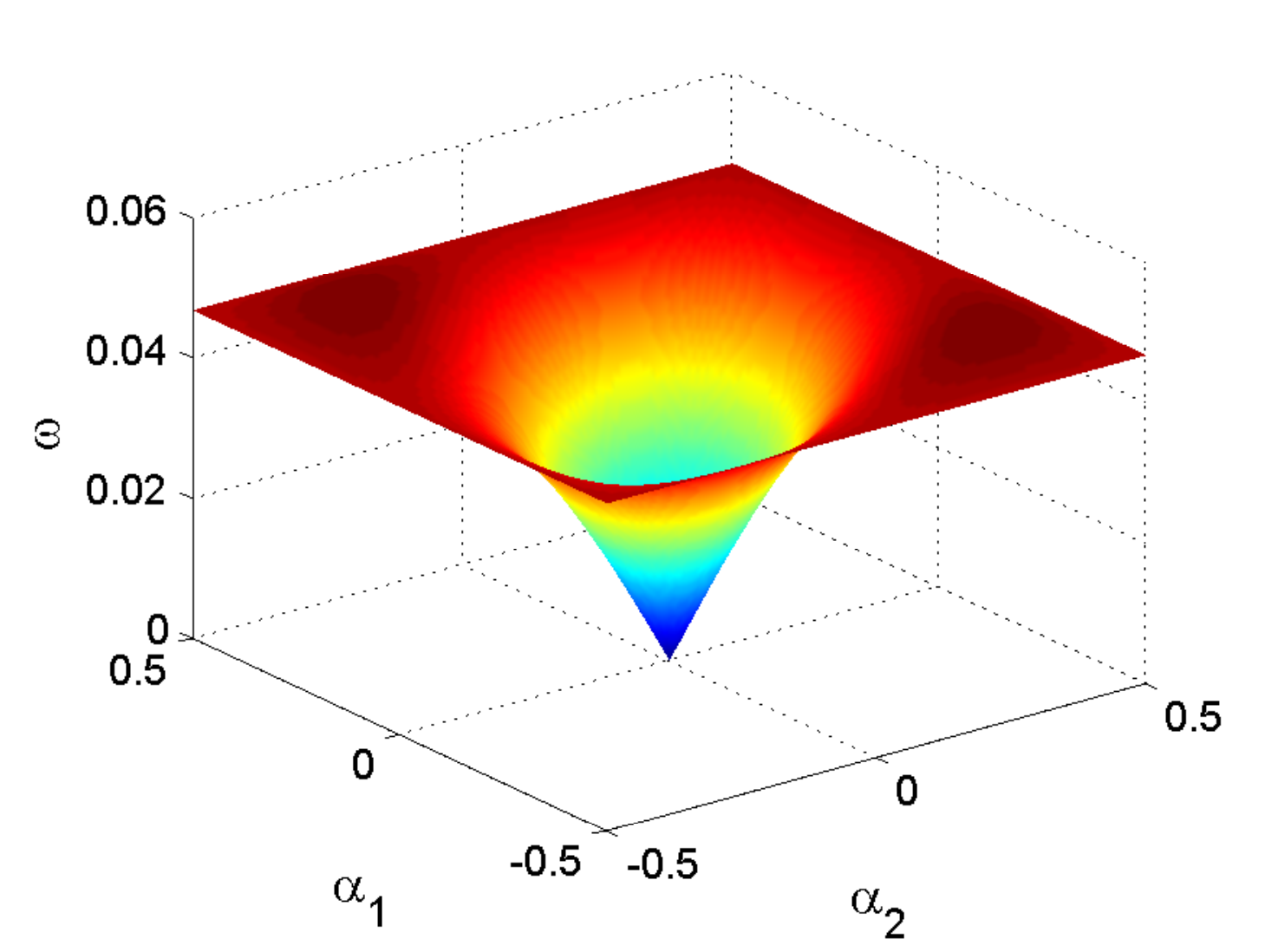}
\caption{(Color online) Bogoliubov excitation spectrum $\omega(\mathbf{k})$. Parameters are 
the same as in Fig.~\ref{fig:phase_QMC}. 
The top panel corresponds to the Mott Phase $\mu/U=0$ and the bottom
one to the superfluid phase $\mu/U=0.15$. As expected, the excitations
are gapped in the Mott phase, whereas, in the superfluid phase, they
are gapless and exhibit a linear behavior at small
momenta.} \label{fig:Bogo_MF}
\end{figure}

As explained above, within the mean-field approach, the Bogoliubov
excitations are obtained from the diagonalization of
$\mathcal{L}_{\mathbf{k}}$, leading to a band structure for the
Bogoliubov spectrum. For instance, in Fig.~\ref{fig:Bogo_MF}, we show
the lowest excitation band for the honeycomb lattice for two different
phases, namely the Mott phase ($\mu/U = 0$) and the superfluid phase
($\mu/U = 0.15$). The rest of the parameters are the same as in
Fig.~\ref{fig:phase_QMC}. As expected, the excitations are gapped in
the Mott phase, whereas, in the superfluid phase, they are gapless and
exhibit a linear behavior at small momenta.

\begin{figure}[h!]
  \centerline{\includegraphics[scale=0.5]{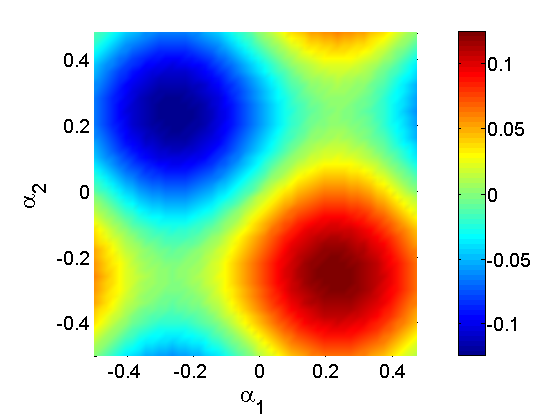}}
  \centerline{\includegraphics[scale=0.5]{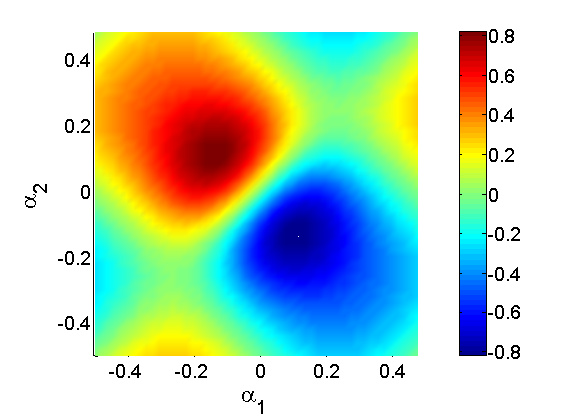}}
\caption{\label{fig:Berry_MF}(Color online) Berry curvature of the Bogoliubov
excitations displayed in Fig.~\ref{fig:Bogo_MF}. 
One can clearly observe $\Omega(\mathbf{k})=-\Omega(-\mathbf{k})$ (time reversal symmetry), 
the antisymmetry with respect to the diagonal ($k_y\rightarrow-k_y$) and 
the symmetry with respect to the anti-diagonal ($k_x\rightarrow-k_x$).}
\end{figure}

In general the Bogoliubov bands are isolated allowing the numerical
computation~\cite{Fukui2005} of the Berry curvature associated with
each band, as shown in Fig.~\ref{fig:Berry_MF}, for the two excitation
bands of Fig.~\ref{fig:Bogo_MF}. One clearly sees that in both
situations, the Berry curvature is non-vanishing, being maximum (in
absolute value) along the anti-diagonal, i.e. the
$k_y$ axis, like in the non-interacting case. In addition, the time
reversal symmetry and the $y\rightarrow-y$ symmetry, like in the non-interacting case, imply that
the Berry curvature is odd under both transformations $\mathbf{k}\rightarrow-\mathbf{k}$ and
$k_y\rightarrow-k_y$ and that it is even under the transformation  $k_x\rightarrow-k_x$.
In  both Fig.~\ref{fig:Berry_MF}  and  Fig.~\ref{fig:Bogo_MF}, these properties
are clearly seen: (i) antisymmetric under inversion and with respect to the diagonal; 
(ii) symmetric with respect to the anti-diagonal.

\begin{figure}[h!]
\centerline{\includegraphics[width=8cm]{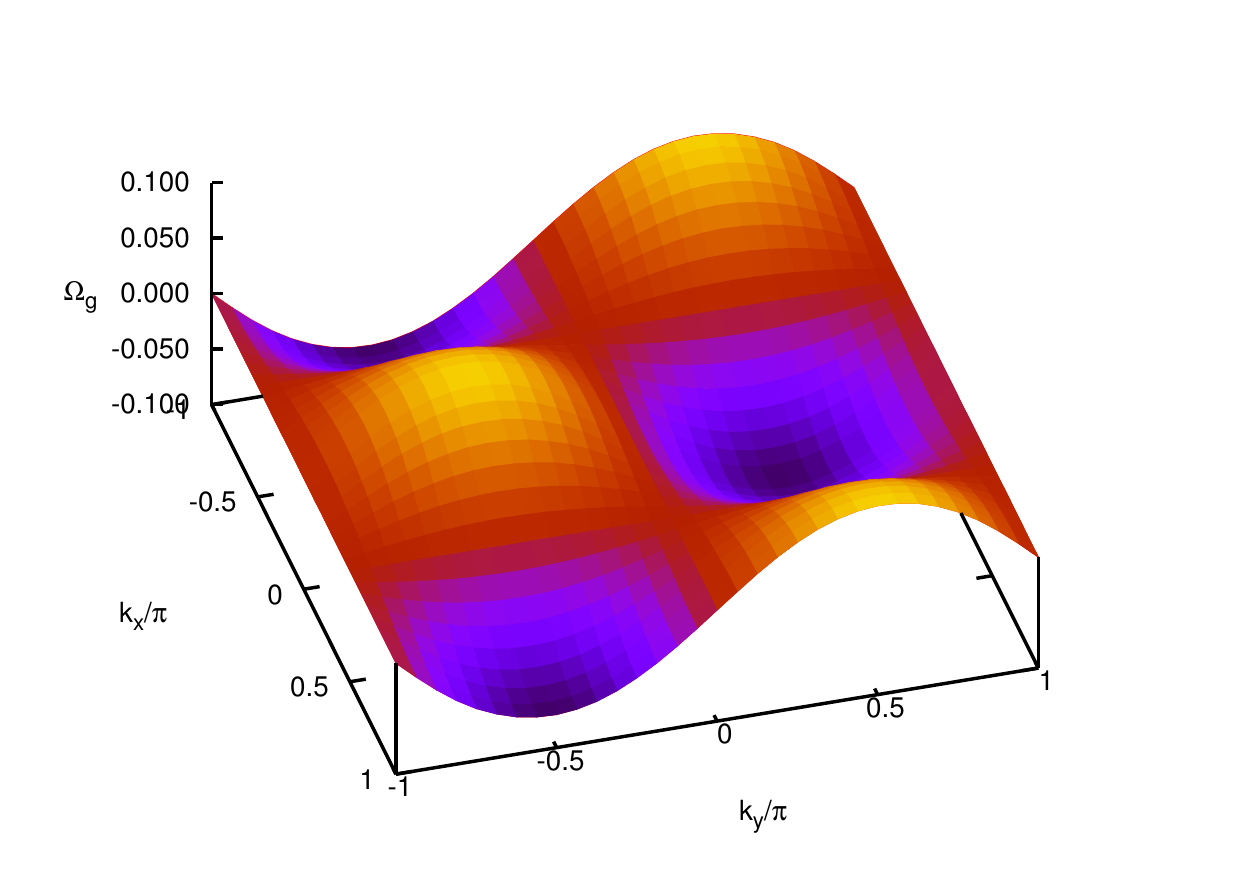}}
\centerline{\includegraphics[width=8cm]{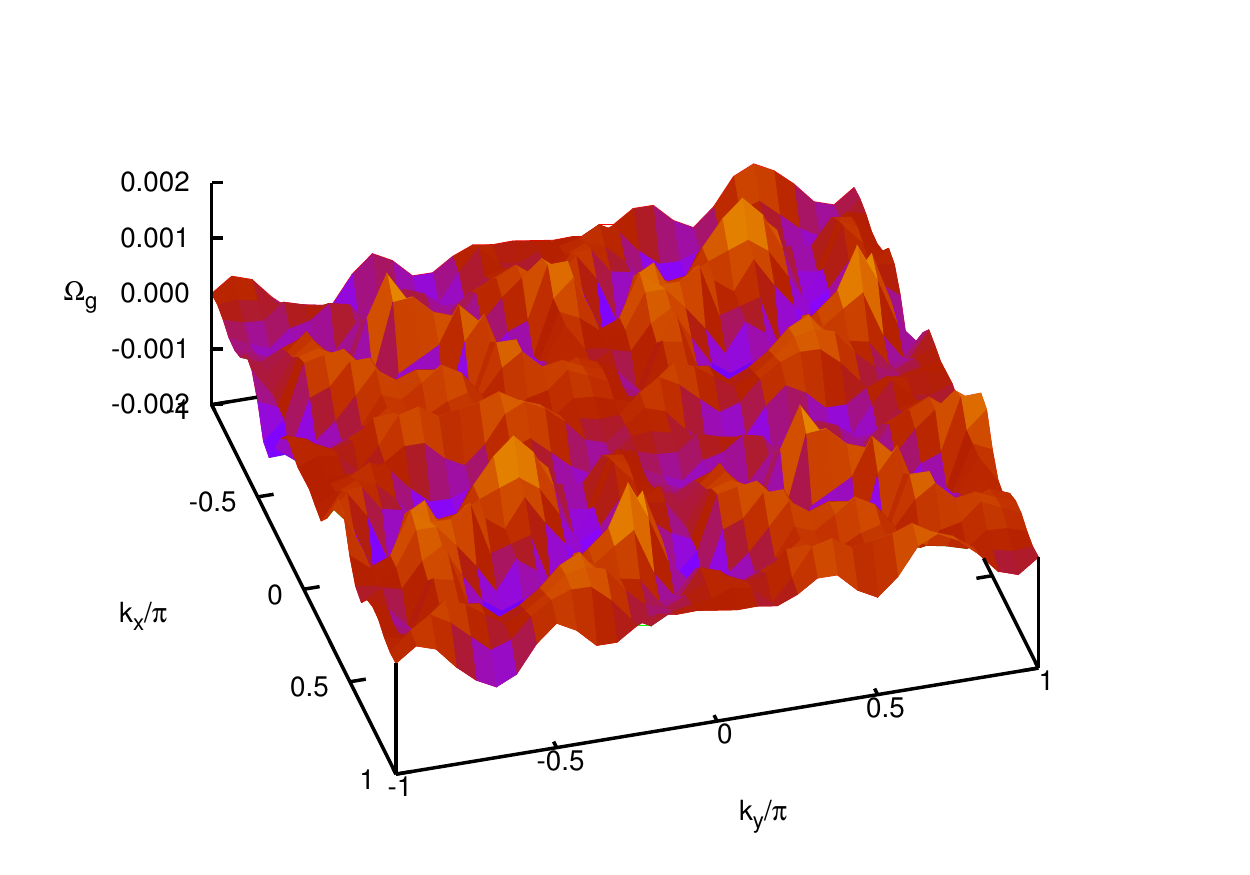}}
\caption{\label{fig:Berry_qmc}(Color online) Pseudo Berry curvature,
  Eq.(\ref{eq:Berry_g}), obtained with the QMC equal-time Green
  functions for the honeycomb and square lattices. The upper panel
  corresponds to the $n=1/2$ Mott phase for the honeycomb lattice,
  i.e. $t_{1,\,2,\,3}=t$, $t_4=0$, $\Delta/U=0.15$, $\mu/U=0$ and
  $t/U=0.03$. The lower panel corresponds to the $n=1/2$ Mott phase
  for the square lattice, i.e. $t_{1,\,2,\,3,\,4}=t$.  For the square
  lattice, one obtains only a random pattern with an amplitude two
  orders of magnitude lower than the honeycomb lattice results. This
  emphasizes that the interacting system must really depict a
  non-vanishing Berry curvature in the excitation spectrum. In
  addition, one can see that the pseudo curvature has  the same
  properties as the non-interacting system: it attains the largest
  (absolute) values around the location of the conical intersections and fulfills
  $\Omega_g(\mathbf{k})=-\Omega_g(-\mathbf{k})$, due to the time reversal
  symmetry. It is also odd under $k_y\rightarrow-k_y$ and 
even under $k_x\rightarrow-k_x$.  }
\end{figure}

Similarly, one could define a pseudo Berry curvature for the
excitations obtained from QMC computations: because of the bipartite
nature of the honeycomb lattice, the equal time Green function in
$\mathbf{k}$-space is a $2\times2$ matrix:
\begin{equation}
 \mathbf{G}(\mathbf{k})=
 \left(\begin{array}{cc}
 G_{AA}(\mathbf{k}) & G_{AB}(\mathbf{k}) \\
 G_{BA}(\mathbf{k}) & G_{BB}(\mathbf{k})
 \end{array}\right),
\end{equation}
where $G_{LL'}(\mathbf{k})=\langle b^{\dagger}_{\mathbf{k}L}
b_{\mathbf{k}L'}\rangle$. Since $\mathbf{G}(\mathbf{k})$ is
hermitian, one can write $\mathbf{G}(\mathbf{k})= g_0(\mathbf{k})
\openone+\mathbf{g}(\mathbf{k})\cdot\bm{\sigma}$, where
$\bm{\sigma}=(\sigma_x,\sigma_y,\sigma_z)$ are the Pauli matrices.
Thereby, one can define the spherical angles $\beta_g(\mathbf{k})$ and
$\phi_g(\mathbf{k})$, corresponding to the direction of
$\mathbf{g}(\mathbf{k})$, or, equivalently, obtained from the
diagonalization of $\mathbf{G}(\mathbf{k})$, similarly to
diagonalizing $H_0$, see Eq.\eqref{eq:Ham_k} and
Eq.\eqref{eq:eigenfunctions_k}. From these two angles, one can define
a pseudo Berry curvature:
\begin{equation}
\Omega_{g}(\mathbf{k}) = \frac{1}{2}\left(
    \frac{\partial\cos\beta_g}{\partial k_x}\frac{\partial\phi_g}{\partial k_y}-
    \frac{\partial\cos\beta_g}{\partial k_y}\frac{\partial\phi_g}{\partial k_x}\right).
\label{eq:Berry_g}
\end{equation}
This curvature, being computed from the equal-time Green function, is
not directly related to a genuine Berry phase that could be measured
from the excitations of the system. For that, one would need to define
a curvature from the real frequency Green function
$\mathbf{G}(\mathbf{k},\omega)$.  That entails computing the imaginary
time Green function and subsequent analytic continuation to the real
axis, which is well known to be a delicate task. Therefore, in order
to emphasize the impact of the lattice topology on the excitations, we
shall focus on the pseudo Berry curvature defined in
Eq.~\eqref{eq:Berry_g}. The results are shown in
figure~\ref{fig:Berry_qmc} corresponding to a ground state in the
$n=1/2$ Mott phase. For comparison, we also show the pseudo Berry
curvature obtained in the same ground state but for the square
lattice. Only the curvature for the honeycomb lattice is non-vanishing
and is largest (in absolute value)
around the same position as for the non-interacting case, i.e. around
the location of the conical intersections. In addition, this pseudo Berry
curvature has the same symmetry properties: $\Omega_g(\mathbf{k})=-\Omega_g(-\mathbf{k})$,
(time reversal symmetry), odd under $k_y\rightarrow-k_y$ and 
even under $k_x\rightarrow-k_x$.

In the superfluid phase, the $2\times2$ matrix is dominated by its
diagonal elements and the resulting pseudo Berry curvature is smaller
than the QMC accuracy.

\afterpage{\clearpage}

\section{Anomalous Hall effect}
\label{dynamics}

Although the preceding analysis has shown that the impact of the Berry
curvature can be found in the properties of the excitations, one can
still raise the question whether it can be observed. It is well known
that in this situation, the system is expected to exhibit an anomalous
Hall effect (AHE) -- a Hall effect without applying an external
magnetic field -- in the non-interacting case. In this section, we
shall demonstrate that one can still observe the AHE in the presence
of the interaction.  However, the AHE can be observed only if the
initial state is a wavepacket well-localized in real space. Indeed, if
one starts with a pure Bloch state, the effect of a constant force only
leads to Bloch oscillations, i.e. a periodic variation of the
quasi-momentum along the direction of the force, and the Berry
curvature simply results in a modification of the phase accumulated
along the path.

\begin{center}
\begin{figure}[h!]
\centerline{\includegraphics[scale=0.55]{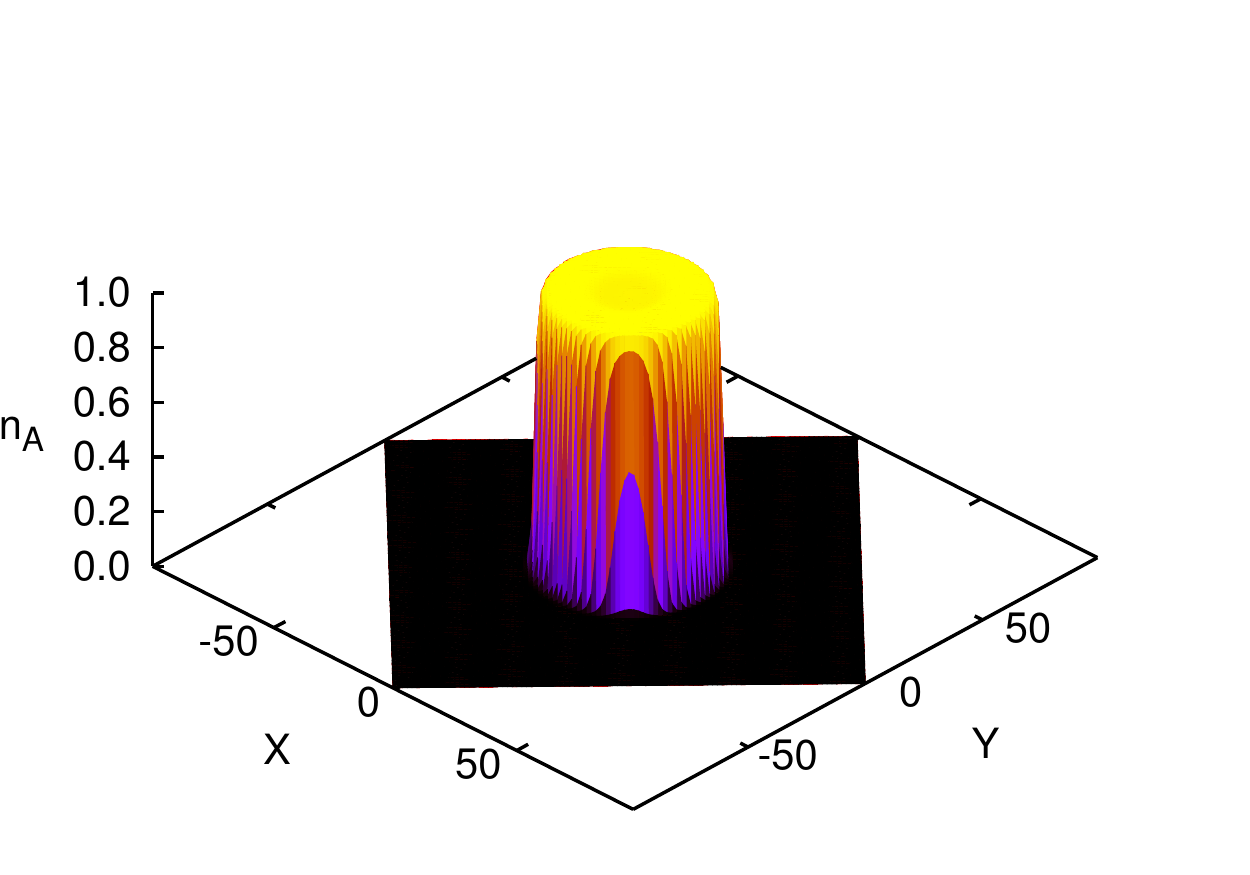}
\includegraphics[scale=0.55]{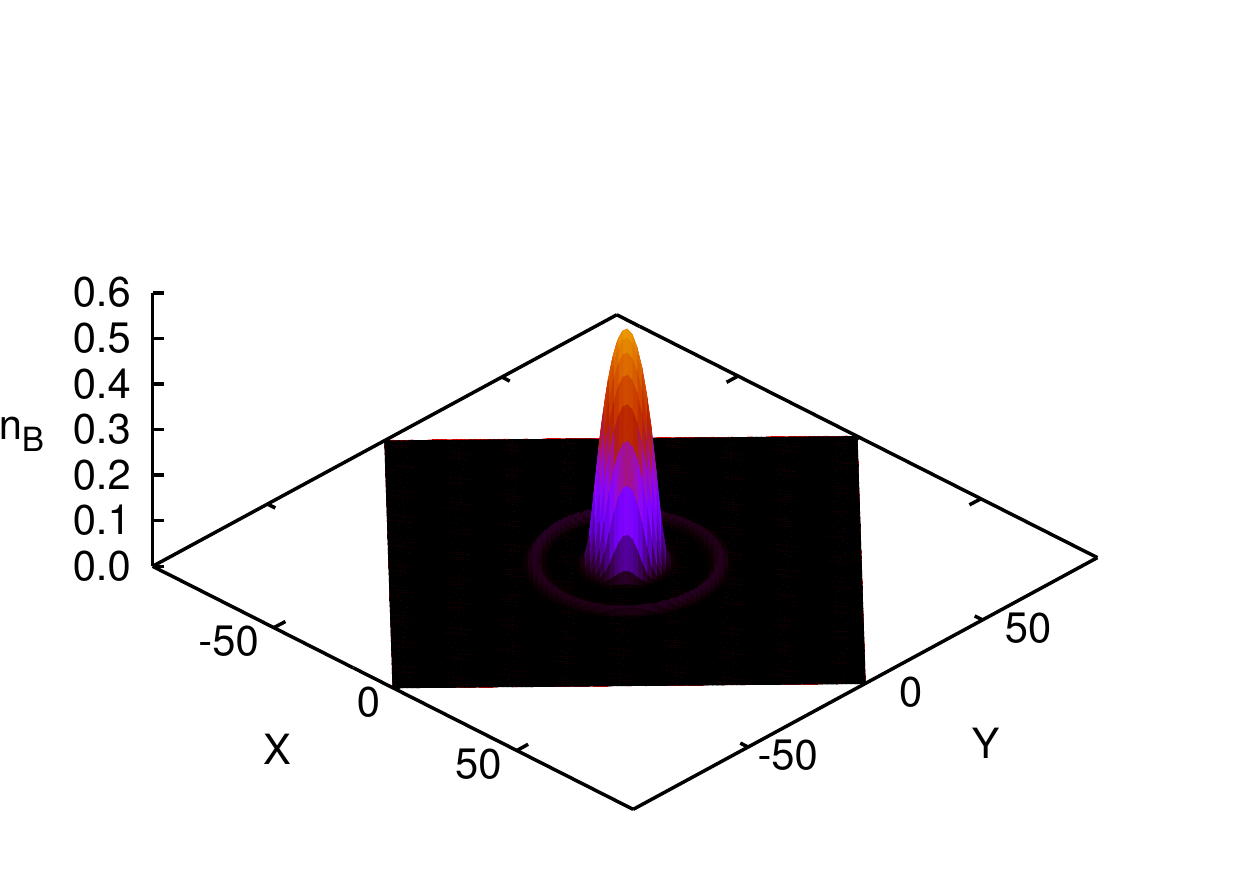}}
\centerline{\includegraphics[scale=0.55]{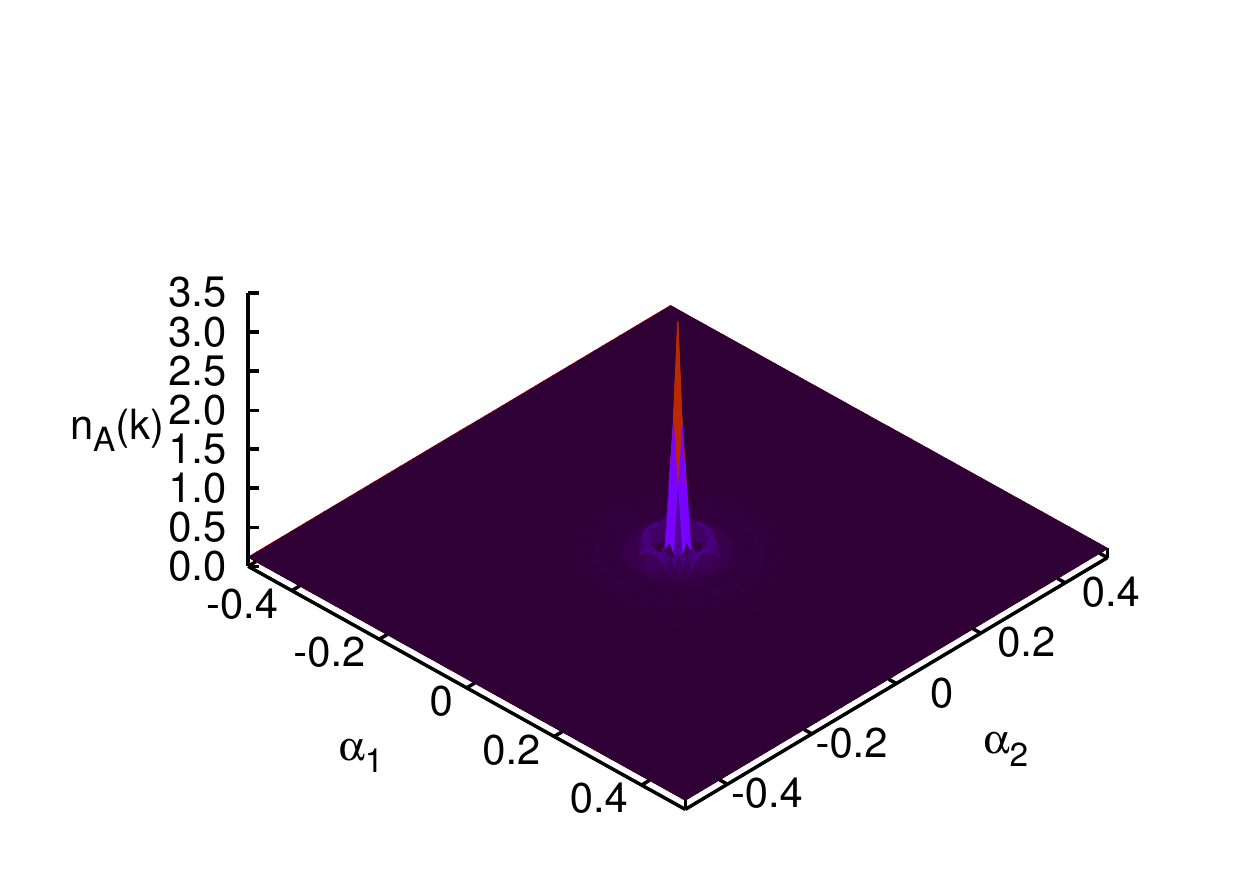}
\includegraphics[scale=0.55]{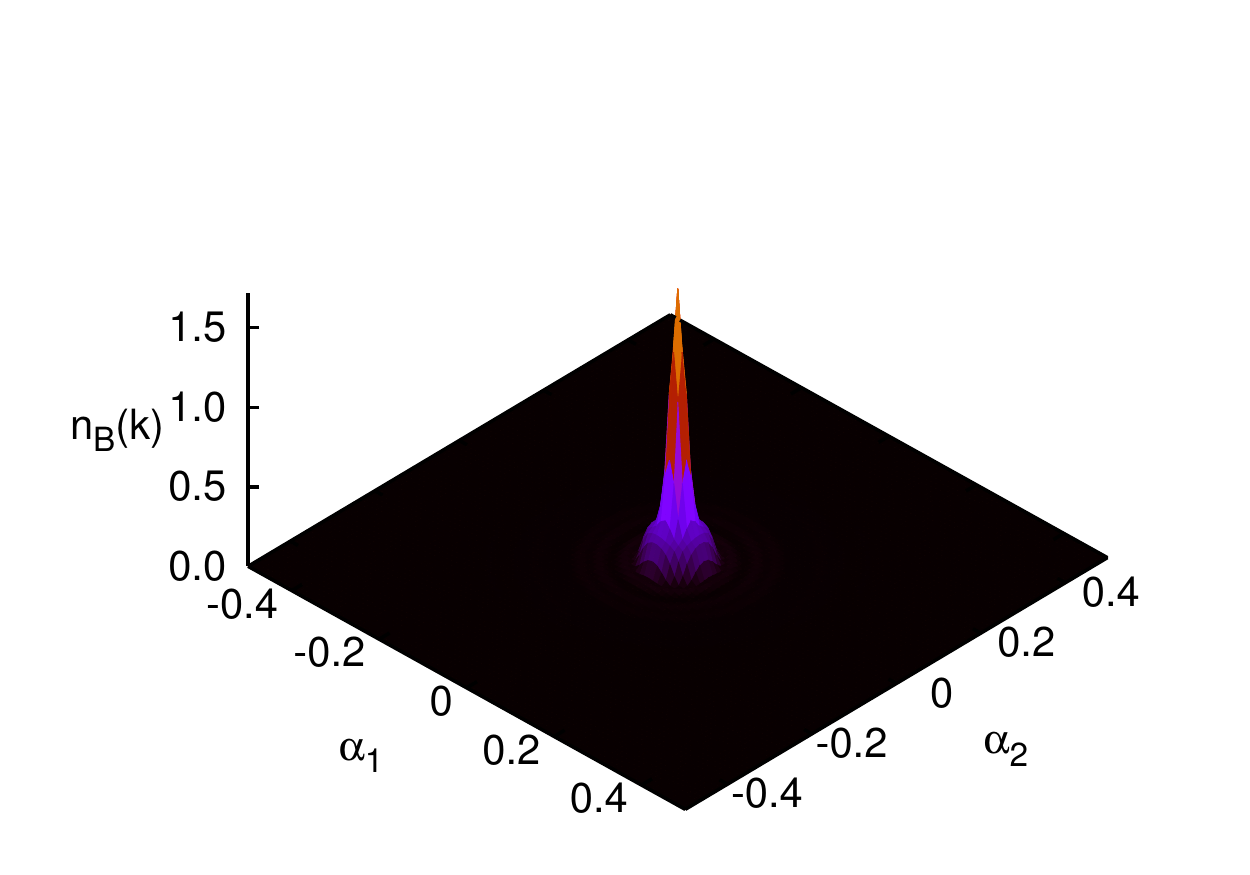}}
\caption{\label{fig:WP_t0}(Color online) Initial wavepacket in a harmonic trap.  The
  atomic density in the $A$-sublattice (resp.  $B$-sublattice) is
  displayed in the top left (resp. top right) plot. The harmonic trap
  is such that the chemical potential at the center corresponds to the
  superfluid phase described in the previous section. Since the
  chemical potential $\mu_A$ is larger than $\mu_B$,
  i.e. $\Delta/U=-0.15$, the $A$-sites are in a Mott-like state with
  unit filling. On the contrary, the filling of the $B$-sublattice is
  low, such that the system is in a superfluid phase. Densities in the
  $\mathbf{k}$-space for both the sublattices are displayed in the
  bottom row and are found to be peaked around $\mathbf{k}=0$. }
\end{figure}
\end{center}

A simple way to prepare a wavepacket is to start from the ground state
in a harmonic trap, as depicted in Fig.~\ref{fig:WP_t0}: the atoms are
well localized in space, still, the wavepacket is also well localized
in the Brillouin zone around $\mathbf{k}=0$. The harmonic trap is such
that the chemical potential at the center corresponds to the
superfluid phase described in the previous section, i.e. $\mu/U=0.15$
and $t/U=0.03$. The site energy of the $A$-sublattice is lower than
that of the $B$-sublattice, corresponding to $\Delta/U=-0.15$,
resulting, in the present case, in a Mott-like state for the $A$-site
with a flat density at unit filling, see Fig.\ref{fig:WP_t0},
top-left. On the contrary, the filling of the $B$-sublattice is low,
such that the system is in a superfluid phase, with a smooth, gaussian
like, density, see Fig.\ref{fig:WP_t0}, top-right. In both cases, the
density in the $\mathbf{k}$-space is peaked at $\mathbf{k}=0$.

\onecolumngrid
\twocolumngrid

\subsection{Bloch oscillations}

\begin{figure}[h!]
\centerline{\includegraphics[scale=0.55]{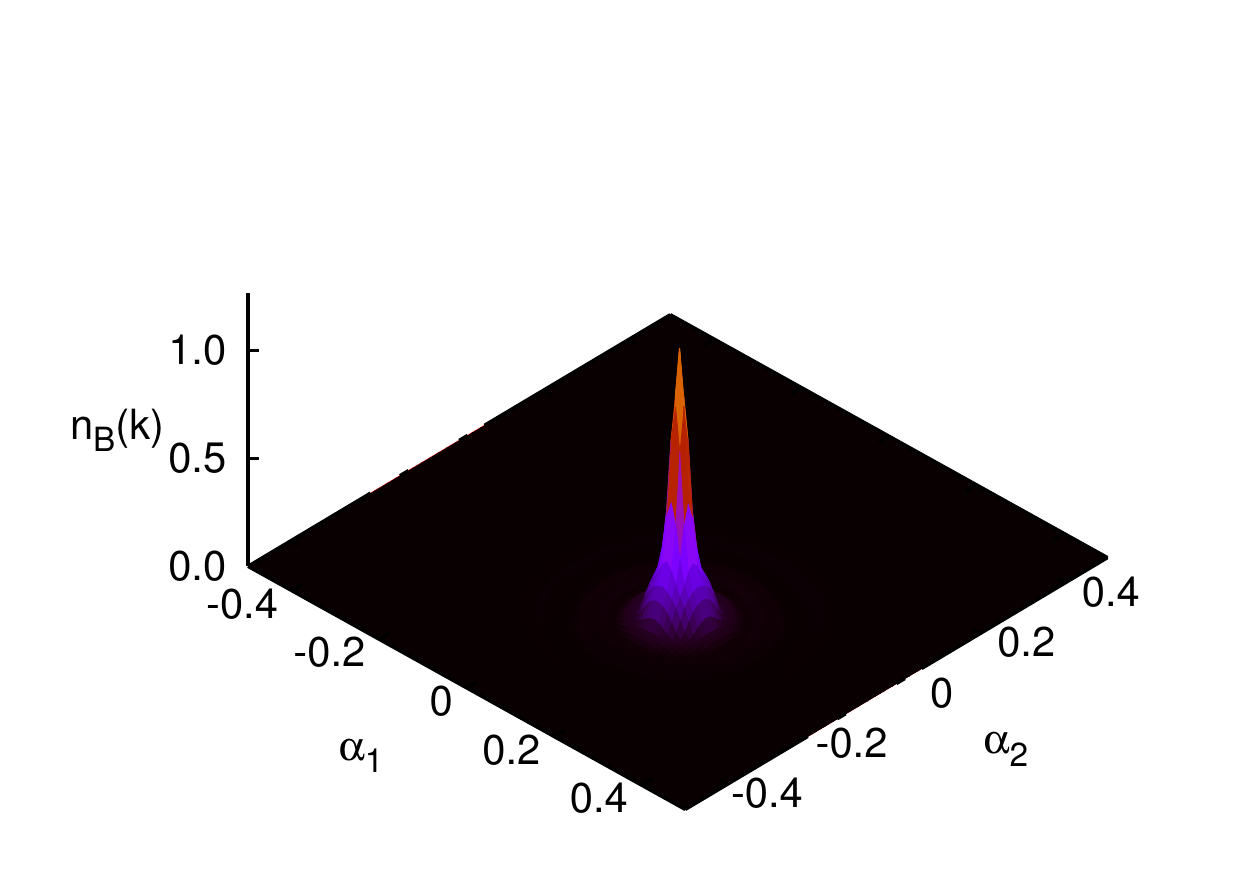}}
\centerline{\includegraphics[scale=0.55]{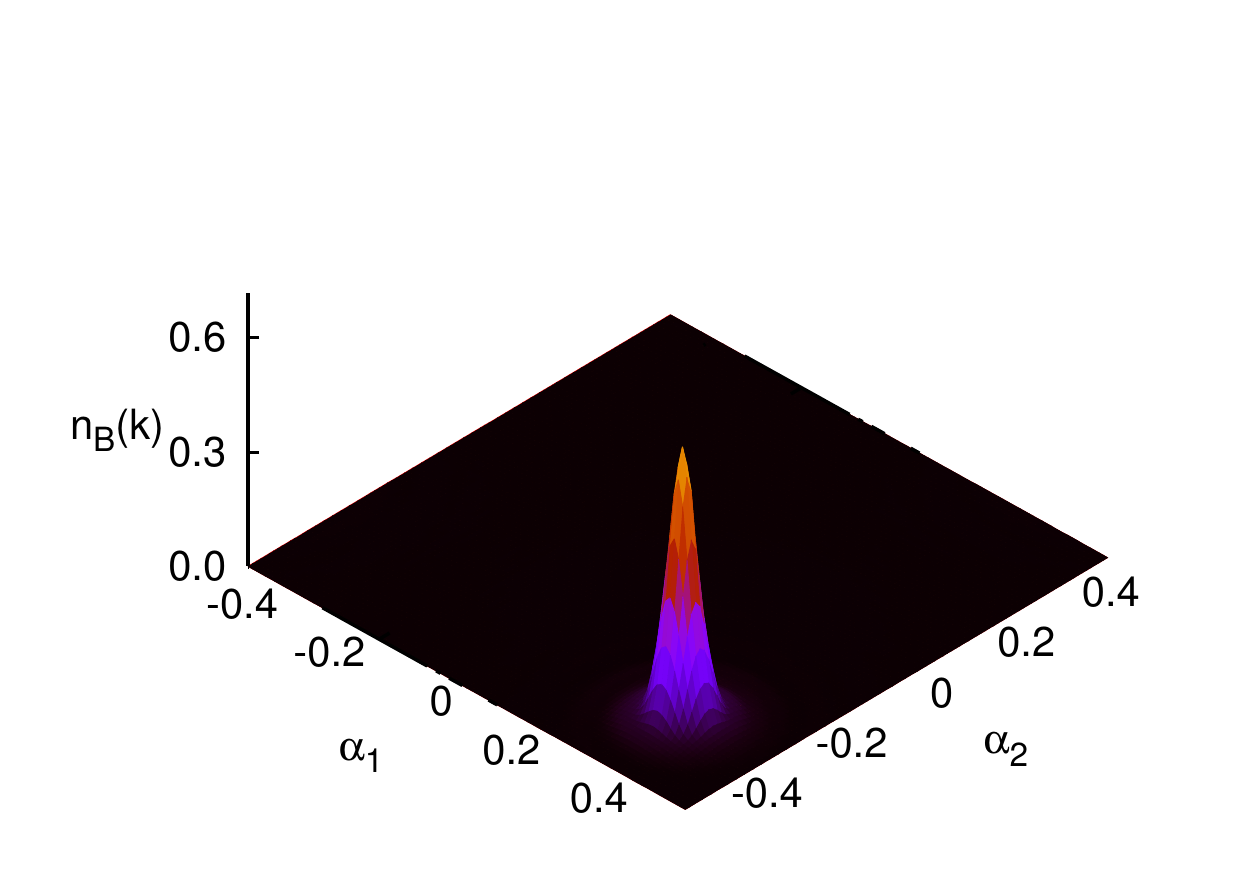}}
\caption{\label{fig:WP_t25_t50}(Color online) Evolution of the wavepacket in
  momentum space. The upper panel displays the wavepacket after
  a time $\tau=25/U$, the lower one after $\tau=50/U$. The center of the wavepacket
  moves at a constant velocity along the $k_y$ axis, i.e.,
  $\alpha_1+\alpha_2=0$. The broadening of the wavepacket is
  weak. }
\end{figure}

\begin{figure}[h!]
\centerline{\includegraphics[width=7cm]{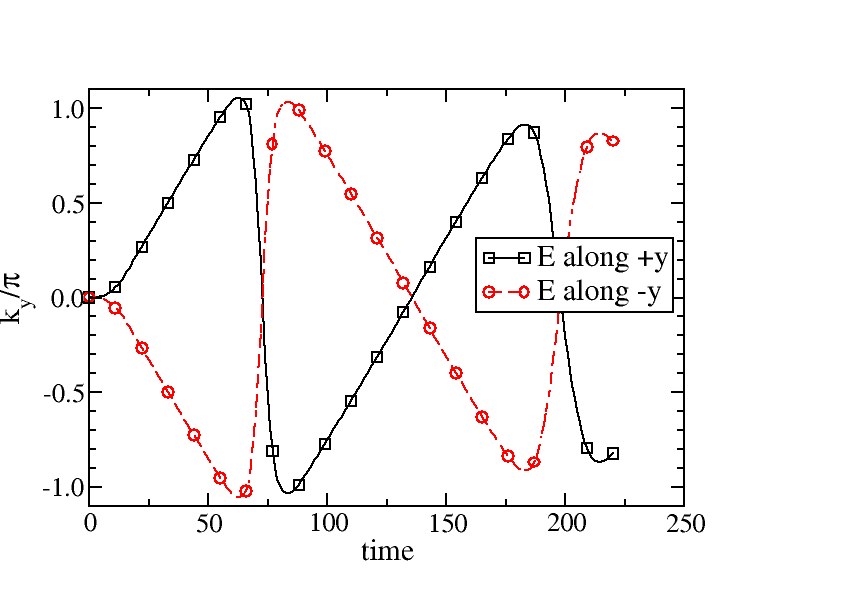}}
\centerline{\includegraphics[width=7cm]{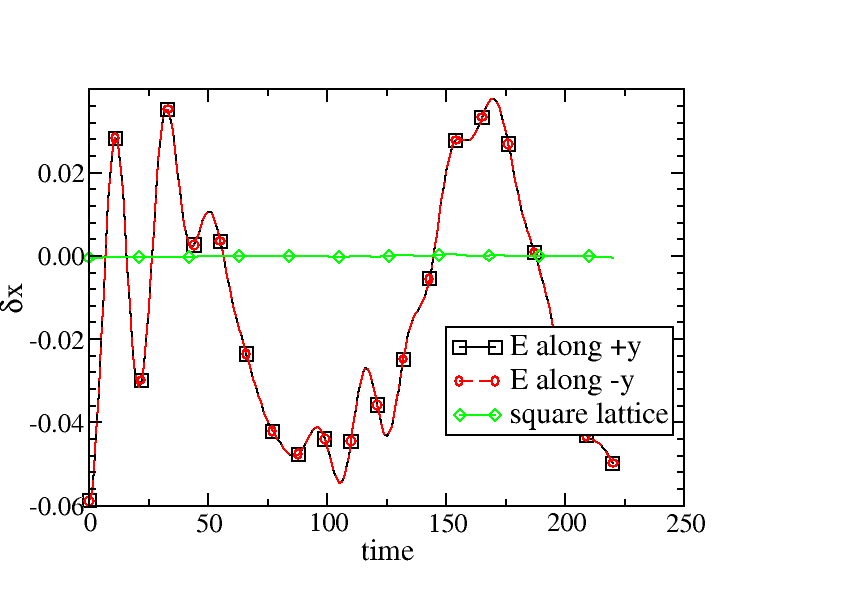}}
\caption{\label{fig:WP_BO}(Color online) Bloch oscillations and AHE for a
  wavepacket. The top plot shows the average value of $k_y$ (within
  the $B$-sublattice) as a function of time. After the adiabatic
  transition, the linear increase of $k_y$ with time is a clear
  signature of the Bloch oscillation. The two curves, i.e, the continuous (black) line 
  and the dashed (red) line with the circles, correspond to
  opposite directions of the effective electric field.  The AHE is
  demonstrated by the bottom plot, where one plots the displacement
  from the center of the trap of the wavepacket along the $x$-axis,
  i.e. perpendicular to the applied force: the continuous (black) line with the squares 
  and the dashed (red) line with the circles correspond to opposite directions of the effective electric field.  For
  comparison, the corresponding displacement on a square lattice is
  also shown by the continuous (green) line with the diamonds.}
\end{figure}

To simulate Bloch oscillations, starting from the wavepacket, one
removes abruptly the harmonic trap and adiabatically increases the
linear potential along the $y$ axis at time $\tau=0$ mimicking the effect of
an electric field $E$ along the $y$ axis. The evolution of the
wavepacket is obtained by solving the time-dependent equations for the
mean-field amplitudes Eq.\eqref{eq:mf_dynamics}. In
$\mathbf{k}$-space, this leads to a smooth evolution of the wavepacket
(with some broadening) along $k_y$, see
Fig.~\ref{fig:WP_t25_t50}. This is emphasized in Fig.~\ref{fig:WP_BO},
where we have plotted the average value of $k_y$ (within the
$B$-sublattice) as as function of time. After the adiabatic
transition, the linear increase of $k_y$ with time is a clear
signature of the Bloch oscillation. The AHE is demonstrated by the
bottom plot in Fig.~\ref{fig:WP_BO}, where we show the displacement
from the center of the trap of the wavepacket along the $x$-axis,
i.e. perpendicular to the applied force. One can clearly see the
impact of the Berry curvature, in particular when comparing with a
similar evolution but for the square lattice, for which there is no
displacement. In addition, changing the sign of the linear potential
i.e. the direction of the applied force, results in changing the
direction of the Bloch oscillations
$\mathbf{k}(\tau)\rightarrow-\mathbf{k}(\tau)$, i.e. changing the direction
of the group velocity of the wavepacket. Together with the fact that
the Berry curvature is odd with respect to the change
$\mathbf{k}\rightarrow-\mathbf{k}$, this results in a Hall
displacement in the \textit{same} direction along the $x$-axis, as
depicted in Fig.~\ref{fig:WP_BO}: the two curves coincide with each
other. Note that we have checked that this effect is independent of
the exact location of the center of the trap. Finally, we have also
verified that if the force is applied along the $x$-axis, no AHE is
observed since the Berry curvature vanishes for $k_y=0$.

\subsection{Shift of the trap center}

\begin{figure}[h!]
\centerline{\includegraphics[width=8cm]{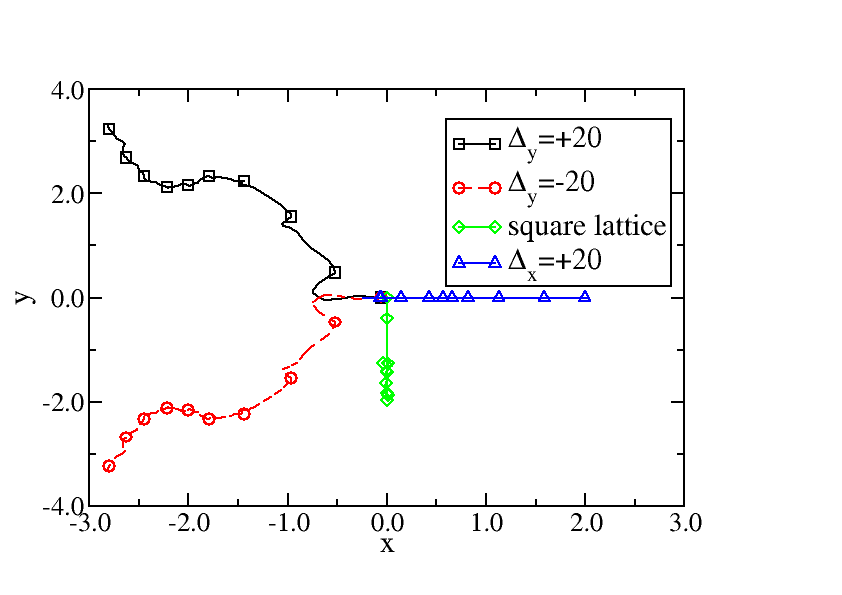}}
\caption{\label{fig:WP_ST}(Color online) Evolution of the average center of the
  wavepacket, i.e., $\langle\mathbf{r}\rangle$, after a shift of the
  trap center by 20 lattice spacings in the $y$ direction (continuous (black)
  line with the squares. The
  wavepacket clearly exhibits motion along the transverse direction,
  i.e. the $x$-axis. Similarly to Bloch oscillations,
  one has the following results: (i) nothing happens on the square
  lattice, see the (green) line with the diamonds; (ii) nothing happens when the shift is along
  the $x$-axis where the Berry curvature always vanishes, see  the (blue) line with the triangles;
  (iii) the displacement along $x$ is in the \textit{same} direction
  whether the trap center is shifted towards the $+y$ or the $-y$
  direction, see the dashed  (red) line with the circles.}
\end{figure}

The amplitude of the AHE discussed above is rather small. One way to
get a larger (measurable) effect consists of shifting abruptly the
center of the trap at $\tau=0$ along the $y$-axis, such that the
wavepacket experiences a net force in that
direction. Figure~\ref{fig:WP_ST} shows the evolution of the
expectation value of the center of the wavepacket, i.e.,
$\langle\mathbf{r}\rangle$, after a shift of the trap center by 20
lattice spacings in the $y$ direction.  As one can see, not only does
the wavepacket evolve along the shifted center, but, at the same time,
it moves along the transverse direction, i.e. the $x$-axis. Similarly
to the Bloch oscillations, one has the following results: (i) nothing
happens on the square lattice; (ii) nothing happens when the shift is
along the $x$-axis where the Berry curvature always vanishes;
(iii) the displacement along $x$ is in the \textit{same} direction
whether the trap center is shifted towards the $+y$ or the $-y$
direction.

\section{Conclusion}
\label{conclusion}
In summary, we have shown that for bosons in the honeycomb lattice
with energy imbalance, the Berry curvature originally seen in the band
structure, i.e. for the non-interacting system, is also present in the
excitations above the ground state of the interacting system, in both
the Mott-insulating phase and the superfluid phase. In addition, we
have shown that one consequence of the Berry phase, the anomalous Hall
effect, could be observed in dynamical experiments, such as
Bloch oscillations.

The Centre for Quantum Technologies is a Research Centre of Excellence
funded by the Ministry of Education and National Research Foundation
of Singapore. PS acknowledges financial support from the Ministry of
Education, Singapore via grant MOE2014-T2-2-112. LY acknowledges
support from the ARC Discovery Projects (Grant Nos DE150101636 and
DP140103231).

\end{document}